\def\address#1{\date{{\sl#1}\\\ \\\version}\gdef\date##1{}}%
\documentclass[twoside,11pt]{article}
\usepackage{cite,amsfonts,a4wide}
\input amssym.def
\input amssym.tex

\newcommand{\cl}[1]{\mbox{${\cal #1}$}}

\newcommand{\BB}[1]{\mbox{${\Bbb #1}$}}

\newcommand {\qd}{\mbox{$\quad$}}
\newcommand {\be}{\begin{equation}}
\newcommand {\e}{\end{equation}}
\newcommand {\bea}{\begin{eqnarray}}
\newcommand {\ea}{\end{eqnarray}}

\newcommand {\bit}{\bibitem}
\newcommand {\bo}[1]{{\bf #1}}

\newcommand {\g}{{\mathfrak g}}
\newcommand {\h}{{\mathfrak h}}
\newcommand {\kf}{{\mathfrak k}}

\newcommand {\al}{\alpha}
\newcommand {\ba}{\beta}
\newcommand {\ga}{\gamma}
\newcommand {\Ga}{\Gamma}
\newcommand {\da}{\delta}

\newcommand {\la}{\lambda}
\newcommand {\La}{\Lambda}
\newcommand {\eps}{\epsilon}

\newcommand {\veps}{\varepsilon}

\newcommand{\frack}[2]{\mbox{$\frac{#1}{#2}$}}
\newcommand {\fract}[2]{\mbox{${\textstyle{\frac{#1}{#2}}}$}}

\begin{document}

\title{
Lie algebra and invariant tensor technology for $g_2$
 } 
\author{
  A.~J.~Macfarlane\thanks{e-mail: A.J.Macfarlane@damtp.cam.ac.uk}
 }
\address{Centre for Mathematical Sciences,\\
  Department of Applied Mathematics and Theoretical Physics,\\
  Wilberforce Road, Cambridge CB3 0WA, UK} 
\maketitle

\begin{abstract}
Proceeding in analogy with $su(n)$ work on $\lambda$ matrices and $f$- 
and $d$-tensors,
this paper develops the technology of the Lie algebra $g_2$, its seven
dimensional defining representation $\ga$ and the full set of invariant
tensors that arise in relation thereto.
A comprehensive listing of identities involving these tensors is given.
This includes identities that depend on use of characteristic equations,
especially for $\ga$, and a good body of results involving the quadratic, 
sextic and (the non-primitivity of) other Casimir operators of $g_2$.

\end{abstract}
\newpage
\tableofcontents
\newpage
\section{Introduction}

This paper is devoted to the detailed study of such aspects of

$\qquad$ a) the Lie algebra $g_2$ with generators $X_i$,

$\qquad$ b) the matrices $x_i$ of its defining $7\times 7$ 
representation $\ga: X_i \mapsto
x_i$,

$\qquad$ c) the $L$-operator $L=x_i \, X_i$,

$\qquad$ d) the $g_2$ invariant tensors that arise in the product laws
associated with the $x_i$, 

\noindent as are expected to be useful in the study of integrable systems for 
which $g_2$ is an invariance algebra.

We begin by giving references that 
have been found useful for general background information on Lie algebras
\cite{hum, FS, corn, FH, slan}, 
and for detailed information
\cite{gt, ram, bdfl, muk, okubo} 
on the exceptional Lie algabra $g_2$.
For some indication of the programme we are aiming to follow for $g_2$, we cite
\cite{mw} 
a study of some integrable systems with the invariance algebra 
$c_n=sp(2n,{\BB R})$. A $g_2$ application resembling the work 
\cite{dVN}
might usefully be undertaken.

Our approach is based on the fact that $g_2$ is a non-symmetric  
subalgebra of $b_3=so(7)$, which is a symmetric subalgebra of $a_6=su(7)$, the
defining representation of all three being of dimension seven.
To explain the distinction made here, let $\g$ be a Lie algebra and $\h$
a Lie subalgebra so that as vector spaces $\g= \h +\kf$. Then in 
addition to
\be \label{0.1} {[} \h \, , \h {]} \subset \h \; , \;
 {[} \h \, , \kf {]} \subset \kf  \qd , \e
we have closure of ${[} \kf \, , \kf {]}$ on $\h$ iff $\h$ is a symmetric 
subalgebra of $\g$, but a more general result 
\be \label{0.2}  {[} \kf \, , \kf {]} \subset \h +\kf \qd, \e
for non-symmetric cases like $g_2=\h \subset \g=b_3$.

Our notation uses indices
\bea
i, j, k \dots & \in & \{ 1,2 \dots ,14 \} \nonumber \\
a, b, c  \dots & \in & \{ 1,2 \dots ,7\} \nonumber \\
\la, \mu, \nu  \dots & \in & \{ 1,2 \dots , 21 \} \nonumber \\
\al, \ba, \ga  \dots & \in & \{ 1,2 \dots ,27 \} \nonumber \\
A, B, C  \dots & \in & \{ 1,2 \dots ,48 \} \qd . \label{0.3} \ea
Thus $X_i, X_a, X_\mu, X_\al, X_A$ respectively 
denote generators of $g_2$, generators of 
$b_3$ which lie outside its $g_2$ subalgebra, generators of $b_3$, generators
of $a_6$ which lie outside its $b_3$ subalgebra, and generators of $a_6$.
Occasionally also we use $A, B, C$ to denote
the generators of an arbitrary simple Lie algebra.

We begin by presenting $b_3$ in Cartan-Weyl form, and $g_2$ as a 
subalgebra of it also in such a form. In the spirt of 
\cite{mw, mpw},
we encapsulate the information 
this entails in terms of the $L$-operators of $b_3$ and 
exhibiting all the matrices of the defining representations $\Ga: X_\mu \mapsto
x_\mu$ of $b_3$, and $\ga: X_i \mapsto x_i$ of $g_2$, simultaneously 
in each case, in a single $7 \times 7$ array. These arrays a) allow convenient
checking of claimed properties like Lie algebra relations, and b) are
valuable artifacts, well-suited for use, {\it cf.} 
\cite{mw},
in the solution of Yang-Baxter
equations of $b_3$ and $g_2$. Likewise matrices such as $A=A_i \, x_i$,
where $A_i \in {\BB R}$ allow easy check of trace and other properties of the 
$x_i$.

To exploit the value of the $L$-operators fully, one needs not only the 
matrices $x_i$ of $g_2$, and the $7 \times 7$ matrices $z_a$ of the defining 
representation of $b_3$ which lie outside $g_2$, but also the matrices $y_\al$,
which along with the $x_i$ and the $z_a$ span the vector space in which the $7 \times 7$ defining representation of $a_6$ acts. 
Detailed information about all these matrices and 
about the various $g_2$ invariant tensors that enter various product laws
(see Sec. 3) must be assembled. To obtain it, it is worthwhile to pass
by similarity transformation
\be \label{0.4} x_i \equiv H_i \; , \; z_a \equiv C_a \; , \;
y_\al \equiv Y_\al \qd , \e
from 
the matrix representions that our Cartan-Weyl starting point naturally 
leads us to, to an equivalent representation in which the matrices enjoy the 
simple transposition properties
\be \label{0.5}
(H_i)^T= -H_i \; ,\; (C_a)^T= -C_a \; , \;
(Y_\al)^T= +Y_\al, \qd . \e The invariant tensors that occur in the product 
laws are of course unaffected by a similarity transformation, and their 
properties are more easily derived. We should remark however that we keep the 
original matrices in practical uses of our $L$-operators. One feature of the 
discussion deserves emphasis. Our product laws in either of their equivalent
forms
\bea 
{[} z_a \, , \, z_b {]} & = & ic_{abc} z_c +ih_{iab} x_i \label{0.6} \\
{[} C_a \, , \, C_b {]} & = & ic_{abc} C_c +ih_{iab} H_i \qd , \label{0.7} \ea
reflect by their first non-trivial first terms the fact that
$g_2$ is not a symmetric subalgebra of $b_3$, and feature the totally 
antisymmetric third
rank quantity $\psi_{abc}$ that occurs in the multiplication law of 
octonions in two roles. These correspond to 
\be \label{0.8} c_{abc}= \sqrt{\fract{1}{3}} \psi_{abc} \qd , \qd 
(C_a)_{bc}= ic_{abc} \qd . \e

We discuss identities including trace and completeness properties of 
matrices, contraction formulas and identities of first class for the
invariant tensors. By first class identities we mean those that stem from 
identities of Jacobi type, ones which apply uniformly throughout say the $a_n$
family, as opposed to the second class identities that involve the use in 
some way of characteristic equations and are specific to each Lie algebra
\cite{msw, mpf}.
To access identities of the second class, we consider characteristic 
polynomials, which also give us a lot of important information about Casimir 
operators. Our approach follows a general method described in generality
elsewhere \cite{mpf} and depends on the algebra of the projectors for the
reduction of the representation $ad \otimes ad$, where $ad$ stands for adjoint,
of $g_2$. Many results emerge like the fact 
\cite{ok2,ok3,mey}
that the quartic Casimir of 
$g_2$ is not primitive, and also formulas for the traces of symmetrised 
products of adjoint matrices of $g_2$. Similar formulas for 
the traces of symmetrised products of matrices $x_i$ can likewise be derived 
from the study of the characteristic equation of $A=A_i \, x_i$. Finally the 
construction, {\it cf.} 
\cite{ok1},
of $g_2$-vectors  and $g_2$-invariants out of the components
$A_i$ of a single adjoint $g_2$ vector is undertaken. We thereby find formulas
expressing naturally occurring non-primitive scalars in terms of suitably
constructed primitive scalars, 
there being one such of each of the orders $2$ and $6$ and no others.

There is of course a very large body of work on $g_2$ already in the 
literature. Our citations have mainly concentrated on works that are directly
to studies and purposes like our own.
However some things
may be given a brief mention to place our work in a wider context.

We point out that $g_2$ has received attention in theoretical physics in
several apparently very different contexts. One was motivated by the search
in progress some forty years ago 
for the flavour symmetry group of the hadrons. This search studied $g_2$ 
and its representations using roots and weights within the Cartan-Weyl
description of Lie algebras. The papers 
\cite{bdfl} \cite{sp} 
are still 
excellent accounts of this work. Many physicist at this time relied on
\cite{racah} for their understanding of Lie algebra theory. 
 
A second definition of $g_2$ is to be found in the work of 
Racah \cite{rac1} (see also \cite{racah}) 
on nuclear spectroscopy. This is couched in terms of
$su(2)$ unit tensors of ranks $1,3$ and $5$ with $3,7$ and $11$ components.
Their commutation relations are determined by
Racah coefficients (or Wigner $6-j$ symbols). 
Not only does the algebra of tensor components close on $b_7=so(7)$, but also, 
because the Racah coefficient
$W(3 \, 5 \, 3 \, 5 \, ; \, 3 \,3)$ vanishes (accidentally ?), the set of $14$
components of the rank one and five unit tensors close on a subalgebra of 
$b_3$, namely $g_2$. The thesis 
\cite{muk} 
contains a clear account of this.
The vanishing of the Racah coefficient is obviously not accidental, since an
analogous view of other exceptional Lie algebras and superalgebras exists,
accessible from 
\cite{vdJ}. See also \cite{vB}.

A third definition of $g_2$ realises it as the subgroup of $b_3=so(7)$ that 
leaves an eight-component $so(7)$ spinor invariant. This is explained in 
\cite{bdfl}
and in many other places; the discussion in \cite{dWn} contains relevant detail
but does not stress it in the $g_2$ context.

A fourth approach to $g_2$ is that of \cite{GMTW}, which expresses the Lie
bracket relations elegantly in terms of generators which transform according to
the octet, triplet and antitriplet representations of its $su(3)$ subalgebra.  

Other matters not considered here (because we deal mainly with certain 
low dimensional ones) include the state labelling problem,
important for discussing the general representation of $g_2$. See {\it e.g.}
\cite{LBR}, where matrix representations are constructed. The problem is 
also fully analysed in \cite{muk}.

While \cite{ok1} discusses the construction of Casimir operators for Lie 
algebras in terms of Lie algebra generators, for $g_2$ an explicit formula 
for the sixth order Casimir is not displayed, although probably accessible. 
See however \cite{amb} 
which describes an explicit construction of the sixth order
Casimir, and a formula for its eigenvalues in a given representation in 
terms of the Dynkin indices of that representation.

See also \cite{gs} (as well as \cite{bdfl}) for information 
about $g_2$ characters and branching rules.

Finally we cite some interesting work on non-compact
realisations of $G_2$ \cite{jos, bnt, bur, sav}.

\section{Cartan-Weyl form of the Lie algebras $b_3$ and $g_2$}
\subsection{General relations}

Let the compact real simple Lie algebra $\g$ with generators $X_A$ and
totally antisymmetric structure constants be defined by
\be \label{1.1} 
[X_A \, , \, X_B] =ic_{ABC} X_C \quad. 
\e
Let $\cl{V}$ denote the defining representation of $\g$ with matrices
$x_A$ given by $X_A \mapsto \gamma x_A$, where $\gamma=1$
for all series of simple Lie algebras  except
$a_n=su(n+1)$, for which $\gamma=\frac{1}{2}$ and $x_A=\la_A$ are a set of 
Gell-Mann matrices 
\cite{msw}. 
We chose the basis so
that
\be \label{1.2} 
  {x_A}^{\dagger}=x_A,\quad {\rm tr}\, x_A=0,\quad
{\rm tr}\,  x_Ax_B=2\delta_{AB}\quad. 
\e
We define also the $L$-matrix by
\be  \label{1.3} 
  L=x_A \otimes X_A \equiv x_A X_A
\e
acting on $\cl{V}\otimes\cl{H}$ where $\cl{V}$ is the defining
representation of $\g$ and $\cl{H}$ any other representation. It is to be
noted that $L$ is not only a quantity of central importance in the study of
integrable systems with invariance algebra $\g$, but that it also encapsulates 
in a very concise and useful manner many properties of $\g$ and its defining 
representation $\cl{V}$.

We present the Lie algebra $\g$ of rank $n$ and $\dim\,\g$ in 
Cartan-Weyl form with generators $\bo{H}
=(H_1,\ldots,H_n)$ of its Cartan subalgebra, positive roots
$\bo{r}_\alpha$, and raising and lowering operators $E_{\pm \alpha}$,
$\alpha \in \{ 1,2,\ldots,\frack{1}{2} (\dim\,\g-n)\}$.  Then we
have
\be \label{2.1}
  [\bo{H},E_{\pm\alpha}] \, = \,  
\pm\bo{r}_\alpha E_{\pm \alpha} \quad , \quad 
  [E_{\alpha},E_{-\alpha}]= \bo{r}_\alpha\cdot\bo{H} \quad, 
\e  
together with well-known non-trivial expressions for 
$[E_{+\alpha} \, , \, E_{\pm \beta} ]$ 
whenever $\bf{r}_{\alpha} \pm \bf{r}_{\beta}$ is
a non-zero root of $\g$. The $X_A$ of (\ref{1.1}) are related to the
Cartan-Weyl generators according to
\be \label{2.2}
   \Bigl\{  X_A\colon A \in\{1,\ldots,\dim\g\}\Bigr\} 
    = \Bigl\{ \begin{array}{l}  H_r \, \colon r \in \{1,\ldots,n\} \\
          U_\alpha \, , \, V_\alpha \, \colon \alpha \in \{1,\ldots,
       \frack{1}{2}(\dim\,\g-n)\} \end{array} \Bigr. \quad  ,
\e
where $\sqrt{2} E_{\pm \alpha}=U_\alpha\pm i V_\alpha $. For the
defining representation $\cl{V}$ of $\g$, we have $X_A \mapsto
x_A$, except, as noted, in the case of $a_n$. Ignoring this case, we employ 
the notation
\be \label{2.3} 
  \bo{H}\mapsto\bo{h}, \quad E_{\pm \alpha}\mapsto e_{\pm \alpha},\quad
  \sqrt{2 }e_{\pm \alpha}= u_{\alpha} \pm i v_{\alpha}\quad , 
\e so that also
\be \label{2.100}
   \Bigl\{  x_A\colon A \in\{1,\ldots,\dim\g\}\Bigr\} 
    = \Bigl\{ \begin{array}{l}  h_r \, \colon r \in \{1,\ldots,n\} \\
          u_\alpha \, , \, v_\alpha \, \colon \alpha \in \{1,\ldots,
       \frack{1}{2}(\dim\,\g-n)\} \end{array} \Bigr. \quad  .
\e

General references for background on Lie algebras and their
representations have been noted 
\cite{hum,FS,corn,FH} as well as the
source 
\cite{slan} of valuable information.

\subsection{The Lie algebra $b_3$}

We turn now to the cases on which the present work focusses.
Thus we start with the Lie algebra of $b_3 \cong so(7)$, 
because we intend to realise
the Lie algebra of $g_2$ as a subalgebra of $b_3$ that is not symmetric.
To distinguish between quantities referring to $b_3$ and their counterparts for
$g_2$, we shall, for $b_3$, use $\bo{R}_\al$ for its roots, $\cl{H}, \cl{E}$ 
for its generators, and $\cl{H} \mapsto k$ and $\cl{E} \mapsto  \eps$ for the 
matrices of the defining representation $\Ga$.

For $b_3$, the simple roots are
\cite{hum}
\be \label{2.21} \bo{R}_1=(1,-1,0) \; , \; 
\bo{R}_2=(0,1,-1) \; , \;
\bo{R}_3=(0,0,1) \quad , \e
and the remaining positive roots are given by
\bea \label{2.22}
\bo{R}_{12}=\bo{R}_{1}+\bo{R}_{2} &=& (1,0,-1)  \quad , \nonumber \\
\bo{R}_{23}=\bo{R}_{2}+\bo{R}_{3} &=& (0,1,0)  \quad , \nonumber \\
\bo{R}_{123}=\bo{R}_{12}+\bo{R}_{3}=\bo{R}_{1}+\bo{R}_{2}+\bo{R}_3 &=& 
(1,0,0)  \quad , \nonumber \\
\bo{R}_{233}=\bo{R}_{23}+\bo{R}_{3}=\bo{R}_{2}+2\bo{R}_3 &=& 
(0,1,1)  \quad , \nonumber \\
\bo{R}_{1233}=\bo{R}_{123}+\bo{R}_{3}=\bo{R}_1+\bo{R}_{2}+2\bo{R}_3 &=& 
(1,0,1)  \quad , \nonumber \\
\bo{R}_{12233}=\bo{R}_{1233}+\bo{R}_{2}=\bo{R}_1+2\bo{R}_{2}+2\bo{R}_3 &=& 
(1,1,0) \quad . \ea
Of these, only $\bo{R}_3, \bo{R}_{23}$ and $\bo{R}_{123}$ are short.

For $b_3$ and $g_2$ it is convenient to denote by $\Gamma$ and $\gamma$
their $7 \times 7$ defining representations $\cl{V}$. We build the matrix
\be \label{2.31}
\cl{L}=
\cl{H}_1\, k_1+\cl{H}_2\, k_2+\cl{H}_3\, k_3+\sum_{\alpha \ge 0} 
(\cl{E}_{\alpha}\, \eps_{-\alpha}+\cl{E}_{\alpha} \, \eps_{\alpha}) \quad , \e
\noindent explicitly in the form 
\be \label{2.32} L= \left(
\begin{array}{ccccccc} 
\cl{H}_1 &&&&&& \\
\cl{E}_1 & \cl{H}_2 &&&&& \\
\cl{E}_{12} & \cl{E}_2 & \cl{H}_3 &&&& \\
\cl{E}_{123} & \cl{E}_{23} & \cl{E}_3 & 0 &&& \\
\cl{E}_{1233} & \cl{E}_{233} & 0 & -\cl{E}_3 & -\cl{H}_3 && \\
\cl{E}_{12233} & 0  & -\cl{E}_{233} & -\cl{E}_{23} & - \cl{E}_2 & \phantom{x} 
 -\cl{H}_2 & \\
0  & -\cl{E}_{12233} & -\cl{E}_{1233} & -\cl{E}_{123} & -\cl{E}_{12} & 
\phantom{x} - \cl{E}_1 & \phantom{xx} -\cl{H}_1 
\end{array} \right) \quad .
\e Here and in all such cases below,
the upper triangular part of $L$ has been suppressed for ease of 
reading but it can easily be supplied with the aid of hermiticity properties.
One may read off (\ref{2.32}) the explicit forms of the 
matrices $k_1, k_2, k_3$ and $\eps_{\pm\alpha}$ of $\Ga$, and verify that they
obey the Cartan-Weyl Lie algebra relations of $b_3$ that follow from
(\ref{2.1}) when the choice (\ref{2.21}) and (\ref{2.22}) of roots is made.
One may write $\cl{L}=\cl{N}_+ + \cl{H} + \cl{N}_-$ to expose the parts of 
$\cl{L}$ upper-triangular,
diagonal and lower-triangular, which corresponds to writing
$b_3=n_+ +h+ n_-$, where $n_{\pm}$ is the span of the  $\eps_{\pm\alpha}$
and $h$ refers to the Cartan 
sub-algebra. Then calculation of ${[} \cl{H} \, , \,  \cl{N}_{\pm} {]}$ and 
${[} \cl{N}_{+}\, , \, \cl{N}_{-} {]}$ is easily done in each case at a single 
(MAPLE) stroke. 

The matrix $\cl{L}$ has many features of interest.
In the main diagonal, and the 
first superdiagonal above it, one sees the non-zero entries of the 
matrices of the Cartan subalgebra
and of those of the lowering operators that belong to the simple roots of 
$b_3$. The matrices $\eps_{-\alpha}$ for roots of heights greater than one are 
then found by looking at the other superdiagonals successively, all the way up
$\eps_{-12233}$ corresponding to the highest root $\bo{R}_{12233}$ of height 
five. One sees how the expansion of any root of $b_3$ in terms of simple 
roots is reflected in $\cl{L}$. 
For example, there is an entry to
$\eps_{-1233}$ in position $(15)$, since $\cl{E}_{1233}$
is associated with the root
$\bo{R}_{1233}=\bo{R}_1+\bo{R}_2+\bo{R}_3+\bo{R}_3$ whose summands are 
similarly related to the entries
in places $(12), (23), (34), (45)$, and so on. Another point is worth 
emphasising because it facilitates the writing down of the matrix $L$ 
for $g_2$ below: all the entries of the $p$-th subdiagonal 
of (\ref{2.32}) refer to roots
of heights $p, \; p \in \{ 1, \cdots , 5\}$, with $p=1$ for the simple roots.
 
It was emphasised in \cite{mpw} that such a construction as we have just 
described is available not only for $b_n$ for all $n$, but also for the other
classical families of Lie algebra. It was also mentioned there that an 
analogous result can be obtained for $g_2$ -- a matter to which we turn
below. 

It may also be verified explicitly that the matrices $x_\mu$ of the defining
representation $\Ga =\cl{V}$ of $b_3$, namely 
\be \label{2.33} x_\mu= \{ k_1, k_2, k_3, u_{\al}, v_{\al}, \al = 1, 
\cdots ,9 \} \quad , \e \noindent
where $\sqrt{2} \, \eps_{\pm \al}= u_{\al} \pm i v_{\al}$, and 
$\al =1, \dots ,9$
corresponds to the index set
\be \label{2.34} 
\{1, 2, 3, 12, 23, 123, 233, 1233, 12233 \} \quad , \e \noindent
possess the properties (\ref{1.2}). In addition they satisfy the antisymmetry 
properties
\be \label{2.35}
x_\mu{}^T= -M x_\mu M_{}^{-1} \qd ,
\e 
where the matrix $M=M^T=M^{-1}$ has ones down its main antidagonal and zeros
elsewhere.

\subsection{The Lie algebra $g_2$}

In discussing the Lie algebra of $g_2$, we employ the standard notation.
We thus employ for $g_2$ the simple roots
\be \label{3.1}
\bo{r}_1= (\fract{1}{\sqrt{6}}\, , \, -\fract{1}{\sqrt{2}}) \quad , \quad 
\bo{r}_2= (0\, , \, \sqrt{2}) \quad , \e \noindent
so that the other four positive roots may be taken to be
\be \label{3.2}
\bo{r}_{12}=\bo{r}_1+\bo{r}_2\, , \, 
\bo{r}_{112}=2\bo{r}_1+\bo{r}_2\, , \, 
\bo{r}_{1112}=3\bo{r}_1+\bo{r}_2\, , \, 
\bo{r}_{11122}=3\bo{r}_1+2\bo{r}_2\, . 
\e 
We see that $\bo{r}_\al$ for $\al=1, 12, 112$ are short roots with 
norm-squared $\fract{2}{3}$ while the others are long with norm-squared equal
to two. Then the Lie algebra relations for $g_2$ follow from (\ref{2.1})
supplemented by the relations
\bea
{[} E_1 \, , \, E_2 {]} & = & E_{12} \nonumber \\
{[} E_{12} \, , \, E_1 {]} & = & \fract{2}{\sqrt{3}} E_{112} \nonumber \\
{[} E_1 \, , \, E_{122} {]} & = & E_{1112} \nonumber \\
{[} E_2 \, , \, E_{1112} {]} & = & E_{11122} \nonumber \\
{[} E_{112} \, , \, E_{12} {]} & = & E_{11122} \quad . \label{3.3} \ea 

One can give the embedding of this realisation of $g_2$ in the realisation 
of $b_3$ presented in Sec. 2.2 explicitly:  
\bea
H_1 & = & \sqrt{\fract{1}{3}} (2\cl{H}_1 +\cl{H}_2+\cl{H}_3) \nonumber \\
H_2 & = & \sqrt{\fract{1}{2}} (\cl{H}_2-\cl{H}_3) \nonumber \\
E_1 & = & \sqrt{\fract{1}{3}} \cl{E}_{1}+\sqrt{\fract{2}{3}} \cl{E}_{3} 
\nonumber \\
E_2 & = & \cl{E}_{2} \nonumber \\
E_{12} & = & \sqrt{\fract{1}{3}} \cl{E}_{12}-\sqrt{\fract{2}{3}} \cl{E}_{23} 
\nonumber \\
E_{112} & = & \sqrt{\fract{2}{3}} \cl{E}_{123}+\sqrt{\fract{1}{3}} 
\cl{E}_{233} \nonumber \\
E_{1112} & = & \cl{E}_{1233} \nonumber \\
E_{11122}  & = & \cl{E}_{12233} \qd . \label{3.100} \ea
Again we note the agreement of the heights of the roots on the two 
sides of each of last six equations. 

Turning to the construction of the matrices $x_i$ of the defining 
representation $\gamma$ of $g_2$, we know that it corresponds to the
span of a subset of $14$ of the $21$ matrices taken above for the
defining representation $\Gamma$ of $b_3$. We also know the wieght diagram of
the $7 \times 7$ representation $\gamma$ of $g_2$, which gives us the 
matrices of the Cartan sub-algebra. We expect an expression like that of
(\ref{2.31}) to have the same characteristics for $g_2$ as 
(\ref{2.31}) itself did for $b_3$. Thus one is lead with very liitle difficulty
to the result for $g_2$
\be \label{3.4} L= \left(
\begin{array} {ccccccc}
c H_1  &&&&&& \\
s E_1 & J_1  &&&&& \\
s E_{12} & E_2 & J_2 &&&& \\
c E_{112} & -c  E_{12} & 
c  E_1  & 0 &&& \\
E_{1112} & s E_{112} & 0 &  -c E_1  & 
-J_2  && \\
E_{11122} & 0 & -s E_{112} & c E_{12} & -E_2
& -J_1 & \\
0 & -E_{11122} & -E_{1112} &  -c  E_{112} &
-s E_{12} & -s E_1 & 
-c H_1
\end{array} \right) \quad .
\e 
Here we have used the abbreviations
\bea J_1=(\fract{1}{\sqrt{6}} H_1 +\fract{1}{\sqrt{2}} H_2) \qd & , & \qd
J_2=(\fract{1}{\sqrt{6}} H_1 -\fract{1}{\sqrt{2}}H_2) \nonumber \\
c \equiv \cos \theta =\sqrt{\fract{2}{3}} \qd & , &
s \equiv \sin \theta =\sqrt{\fract{1}{3}} \qd . \label{3.5} \ea

There is little difficulty because $g_2$ has exactly one positive root of each
height $p=2,3,4,5$. Thus working downwards from the highest root, we get
the placing of $E_{11122}$ and $E_{1112}$ directly. For $E_{112}$ we place 
down the third sub-diagonal the entries $(c_1, s_1, -s_1, -c_1)$, with values 
of $c_1$ and $s_1$ to be fixed. Here $c_1=\cos \theta_1$. The entries for 
$E_{12}$ and $E_{1}$ likewsie involve further angles $\theta_2$ and 
$\theta_3$, but the entries for $E_2$ are written down directly.
One reads the implied matrices for the $e_{\pm \al}$ off the display for $L$
partially thereby obtained. Then it is easy to see the assignmnents represent
(\ref{3.3}) correctly for the values presented in (\ref{3.4}) and (\ref{3.5}).
The assignments could also have been inferred in agreement with this using
the defining representation matrices of (\ref{3.100}). 

Some of the useful features of (\ref{3.4}) can be seen more clearly when we 
present in more schematic form with signs and constants supressed
\be \label{3.6} \left(
\begin{array} {ccccccc}
x  &&&&&& \\
1 & x &&&&& \\
12 & 2 & x &&&& \\
112 & 12 & 1 & x &&& \\
1112 & 112 & 0 & 1 & x && \\
11122 & 0 & 112 & 12 & 2 & \phantom{x} x & \\
0 & 11122 & 1112 & 112 & 12 & 1 & \phantom{xx} x
\end{array} \right) \quad .
\e
The pattern of the places associated with the non-simple roots relative to the
simple ones here, despite its non-trivial nature, 
conforms very closely to that found for the $c_n$ and $b_n$
families. See 
\cite{mpw} 
and {\it cf.} (\ref{2.32}).

We may read explicit expressions for the matrices $h_1, h_2$ and $e_{\pm \al}$
from (\ref{3.4}) and (\ref{3.5}), and check that they obey exactly the same 
Lie algebra relations as do the abstract generators. Because the 
$x_i$ for 
$g_2$ are a subset of those given above for $b_3$, they share the same 
properties, namely (\ref{1.2}) and (\ref{2.35}).

\subsection{$g_2$ as a Lie subalgebra of $b_3$}

We have seen that the $14$ matrices $x_i$ of the defining representation $\ga$
of $g_2$ share the same properties as the $21$ matrices of $\Ga$ for $b_3$. 
The explicit
definitions show the former to be simple linear combinations of the latter. It
is natural to want to define the $7$ linear combinations of the $b_3$ matrices
that lie outside $g_2$, but which share the properties (\ref{1.2}) and 
(\ref{2.35}). To do so at a single stroke we define
\be \label{4.1}
K=H_3\, h_3+\sum_{\ba=1}^{3} (E^\prime{}_{\ba}\, e^\prime{}_{-\ba} 
+E^\prime{}_{-\ba}\, e^\prime{}_{\ba}) \quad , \e
\noindent explicitly in the form 
\be \label{4.2} 
K= \left(
\begin{array}{ccccccc} 
s H_3 &&&&&& \\
-c E^\prime{}_{1} & -s H_3 &&&&& \\
c E^\prime{}_{2} & 0 & -s H_3 &&&& \\
s E^\prime{}_{3} &  s E^\prime{}_{2} &  s E^\prime{}_{1} &  0 &&& \\
0 & -c E^\prime{}_{3} & 0 & - s E^\prime{}{_1} & s H_3 && \\
0 & 0  & c E^\prime{}_{3} & - s E^\prime{}_{2} & 0 & \phantom{x}  
s H_3 & \\
0  & 0 & 0 &  -s E^\prime{}_{3} &  -c E^\prime{}_{2} & 
 c E^\prime{}_{1} & \phantom{xx} -s H_3 
\end{array} \right) \quad ,
\e where $c$ and $s$ are as given in (\ref{3.5}).
This implies
\be \label{4.3}
z_4 = h_3= s \, {\rm diag} \, (1, -1, -1, 0, 1, 1, -1) \qd , \e
and we define the other $z_a \; , \; a \in \{ 1, \cdots , \}$ by means of
\bea
\sqrt{2} e_{\pm 1} & = & z_7 \pm i z_1 \nonumber \\
\sqrt{2} e_{\pm 2} & = & z_6 \pm i z_2 \nonumber \\
\sqrt{2} e_{\pm 3} & = & z_5 \pm i z_3 \qd . \label{4.4} \ea
The reason for the arrangement of detail here is provided in Sec. 3.

The display (\ref{4.2}) thus yields explicitly the matrices $z_a, a \in \{ 1, 
\cdots , 7\}$ which span the $b_3 -g_2$ part of the vector space of the 
representation $\Ga$ of $b_3$. The following properties may be checked
\be \label{4.5}
{\rm tr} \, z_a=0 \; , \; {\rm tr} \; (z_a z_b)=\da_{ab} \; , \;  
{\rm tr} \, (z_a x_i)=0 \; , \; z_a{}^{T}=-M z_a M^{-1} \qd . \e
It is to be noted that $K$ simply uses those linear combinations of $b_3$
 matrices that are excluded from (\ref{3.4}).

\subsection{The matrices $y_\al$}

To complete the basis we need for $7 \times 7$ traceless hermitian matrices,
we define a set of $27$  linearly independent matrices $y_\al$ which enjoy
the properties
\be \label{5.1}
y_\al{}^\dagger = y_\al \; , \; {\rm tr}\, y_\al=0 \; , \;
y_\al{}^{T}=+M y_\al M^{-1} \; , \; \al \in \{ 1, \cdots ,27 \} \qd , \e
and 
\be \label{5.2}
{\rm tr}\, (y_\al y_\ba)=2\da_{\al \ba} \; , \; 
{\rm tr}\, (x_i y_\al)=0 \; , \; i \in \{ 1 , \cdots , 14 \} \; , \;
{\rm tr}\, (z_a y_\al)=0 \; , \; a \in \{ 1 , \cdots , 7 \} \; . \e 
By referring to the diagonal matrices of $\ga$ and to the diagonal 
matrix $z_4=h_3$ given by (\ref{4.3}) we chose the three diagonal matrices
$y_\al$ to be 
\bea
y_1 & = &  \fract{1}{\sqrt{6}} {\rm diag} \,(2, -1, -1, 0,- 1, -1, 2) 
\nonumber \\
y_2 & = &  \fract{1}{\sqrt{2}} {\rm diag} \, (0, 1, -1, 0, -1, 1, 0) 
\nonumber \\
y_3 & = &  \fract{1}{\sqrt{21}} {\rm diag} \, (1, 1, 1, -6, 1, 1, 1) 
\quad .\label{5.3} \ea
This completes the choice of a set of six diagonal traceless hermitian 
matrices tracewise mutually orthogonal and also orthogonal to the matrix
${\rm diag} \, (1, 1, 1, 1, 1 ,1, 1)$.
The other $24$ matrices $y_\al$ are easy to specify using the operator matrix
\be \label{5.4}
\sum_{n=1}^{12} (\rho_{-n} R_n + \rho_n R_{-n})
= \left(
\begin{array} {ccccccc}
x  &&&&&& \\
R_1 & x &&&&& \\
R_4 & R_2 & x &&&& \\
R_6 & R_5 & R_3  & x &&& \\
R_8 & R_7 & \sqrt{2} \, R_{12}  & R_3 & x && \\
R_9 & \sqrt{2} \, R_{11} & R_7 & R_5 & R_2 & \phantom{x} x & \\
\sqrt{2} \, R_{10} & R_9 & R_8 & R_6 & R_4 & R_1 & \phantom{xx} x
\end{array} \right) \qd .
\e
The diagonal elements $x$ here can be inferred from (\ref{5.3}).
Then setting
\be \label{5.5}
\sqrt{2} \rho_{\pm n} = y_{2n+2} \pm i  y_{2n+3} \qd , \qd n =1, \cdots ,12
\qd , \e
completes the definition of the $y_\al \, , \, \al \in \{ 1, \dots , 27
\}$. Our display of an explicit choice of matrices $x_i, z_a, y_\al$
is motivated by the need to perform/check our matrix, and later,
tensorial manipulations by MAPLE.

\section{Properties of the $7 \times 7$ matrices of $g_2, b_3$ and $a_6$}

We have defined the traceless hermitian $7 \times 7$ matrices 
of $x_i, i \in  \{ 1, \dots ,14\}$ of $g_2$, 
$z_a, a \in \{ 1, \dots ,7\}$ of $b_3-g_2$, and 
$y_\al, \al \in  \{ 1, \dots ,27 \}$ of $a_6-b_3$. All are normalised so that
the trace of their square is $2$, and they are tracewise mutually orthogonal.
They have the symmetry 
properties
\be \label{10.1}
x_i{}^T= -Mx_i M^{-1} \; ,\;  z_a{}^T= -Mz_a M^{-1} \; , \; 
y_\al{}^T= My_\al M^{-1} \qd . \e
It is possible to view the set of $48$ matrices $x_i, z_a, y_\al$ as a set of
lambda matrices of $a_6$. Their single multiplication rule here is replaced 
by the following
\bea
x_i \, x_j & = & \fract{2}{7} \da_{ij} +\fract{1}{2} i c_{ijk} x_k
+\fract{1}{2} d_{ij\al} y_\al  \label{10.2} \\
z_a \, z_b & = & \fract{2}{7} \da_{ab}+ \fract{1}{2} i c_{abc} z_c
+\fract{1}{2} ih_{iab} x_i +\fract{1}{2} d_{ab\ga} y_\ga \label{10.3} \\
y_\al \, y_\ba & = & \fract{2}{7} \da_{\al \ba}
+\fract{1}{2} i \phi_{i \al \ba} x_i +\fract{1}{2} i t _{a \al \ba} z_a
+\fract{1}{2} d_{\al \ba \ga} y_\ga \label{10.4} \\
x_i \, z_a & = & \fract{1}{2} ih_{iab} z_b+ \fract{1}{2} d_{ia\al} y_\al
\label{10.5} \\
x_i \, y_\al & = & \fract{1}{2} i \phi_{i \al \ba}y_\ba 
+ \fract{1}{2} d_{ij\al} x_j
+ \fract{1}{2} d_{ia\al} z_a \label{10.6} \\
z_a \, y_\al & = & \fract{1}{2} i t _{a \al \ba}y_\ba+\fract{1}{2} d_{ab\al} 
z_b +\fract{1}{2} d_{ia\al} x_i \qd . \label{10.7} \ea
 
First we note that since all tensors here are real, hermitian conjugation of
(\ref{10.2}) -- (\ref{10.4}) implies certain evident symmetry and 
antisymmetry properties.
Second we note that behaviour under conjugation with $M$ is what determines 
which terms 
are allowed on the right hand sides. Third we note 
the tracelessness of symmetric tensors occurring in 
(\ref{10.2}) -- (\ref{10.4})
\be \label{10.100}
d_{ii\al}=0 \; , \; d_{ab\al}=0 \; , \; d_{\al \al \ga}=0 \qd . \e
Fourth we note the absence from 
(\ref{10.2}) of a term in $z_a$. This simply reflects the closure property
of the Lie
algebra $g_2$:
\be \label{10.8}
{[} x_i \, , \, x_j {]}= i c_{ijk} x_k \quad . \e
Fifth (\ref{10.2}) and (\ref{10.4}) imply
\bea
{[} z_a \, , \, z_b {]} & = & i c_{abc} z_c +ih_{iab}x_i \label{10.9A} \\
{[} x_i \, , \, z_a {]} & = &  ih_{iab}z_b \qd . \label{10.9B} \ea
Eqs. (\ref{10.8}), (\ref{10.9A})  and (\ref{10.9B})
represent the Lie algebra relations of 
$b_3$ as generated by $x_i$ and $z_a$. The presence of the $x_i$ term in 
${[} z_a \, , \, z_b {]}$ shows that $g_2$ is not a symmetric subalgebra of
$b_3$ merely a reductive one. 
Sixth we note that, in the use of eqs. (\ref{10.2}) to (\ref{10.7})
to define the various ($g_2$) invariant tensors there are repetitions. This 
follows from easy trace considerations. The latter also imply the total
antisymmetry of $c_{ijk}$ and $c_{abc}$ and the total symmetry of 
$d_{\al \ba \ga}$.
 
The totally antisymmetric $b_3$ structure constants $c_{abc}$ occurring in 
(\ref{10.9A}) are particularly interesting quantities, because, with due
arrangement of the details, they are constant multiples of the totally
antisymmetric tensors that enter the multplication law of octonions. This
should not be entirely surprising in view of the well-known occurrence of 
octonions in connection with $g_2$ featured in \cite{gt} and in the other 
main $g_2$ sources cited.

It is easy to see this explicitly, and appreciate the arrangement of detail in 
(\ref{4.4}), by showing by direct calculation that 
\be \label{10.10}
c_{abc} =\fract{1}{\sqrt{3}} \,  \psi_{abc} \; , \, a, b, c \in \{1, \dots ,7\} \qd , \e
where the totally antisymmetric octonionic tensor $\psi_{abc}$ has the value 
one for these ordered triples
\be \label{10.12}
(123), (147), (165), (246), (257), (354) , (367) \qd , \e
and no other non-zero components except for those implied by antisymmetry.
This is already a striking result, but it will be
further embellished in Sec. 4.1 below. One consequence of (\ref{10.10}) is that
it greatly helps in the search for identities involving invariant tensors 
just defined, because the many known identities involving the $\psi_{abc}$, see
especially \cite{gt}, give us good control of the $c_{abc}$. Octonionic tensor
identities are reviewed in Sec. 3.1.

\subsection{Identities for the octonionic tensor $\psi_{abc}$}

The triples for which the totally antisymmetric third rank tensor
$\psi_{abc}$ takes non-zero values appear in (\ref{10.12}). We note the
following identities which it satisfies
\be \label{A1.1}
\psi_{abc} \, \psi_{abd}=6\da_{cd} \qd . \e
\be \label{A1.2}
\psi_{abc}  = - \frac{1}{4!} \veps_{abcdefg}\psi_{deh} \psi_{fgh}  \qd . \e
Here we see the totally antisymmetric seventh rank epsilon tensor. It is easy 
to confirm the truth of these results directly, and to see that they imply the
key result
\be \label{A1.3}
\psi_{deh} \, \psi_{fgh}=\da_{df} \da_{eg}-\da_{dg} \da_{ef}
-\fract{1}{6} \veps_{defgabc} \psi_{abc} \qd ,  \e
whence we can obtain the results
\bea 
\psi_{fag} \psi_{gbe} \psi_{ecf} & = & 3 \psi_{abc} \label{A1.4} \\
\psi_{h[de} \psi_{f]gh} & = & -\fract{1}{6}\veps_{abcdefg} 
\psi_{abc} \label{A1.5} \\
\psi_{deh} \psi_{fgh} +\psi_{feh} \psi_{dgh} & = & 2\da_{df}\da_{eg}- 
\da_{dg}\da_{ef}-\da_{fg}\da_{ed} \qd . \label{A1.6} \ea

The identity (\ref{A1.3}) is a very simple and convenient product law, but it 
involves the seventh rank epsilon tensor. If one wants a product law that
does not then one must have recourse to the remarkable result
-- derived without use of (\ref{A1.3}) -- from 
\cite{bdfl}
\be \label{A1.7}
\psi_{abd} \psi_{deh} \psi_{hfg}= 3\da_{a[e} \psi_{fg]b}-
3\da_{b[e} \psi_{fg]a}
-\da_{ef} \psi_{gab}+\da_{eg} \psi_{fab} \qd . \e
To establish this ,here on the basis of (\ref{A1.3}), 
we multiply (\ref{A1.3}) by $\psi_{abd}$, and use a simple 
argument to treat the $\eps \da \da$ term that arises.

The result (\ref{10.10}) relates the octonionic tensor $\psi_{abc}$ to 
structure constants $c_{abc}$ of $b_3$. However (\ref{A1.5}) proves that the 
$c_{abc}$ 
{\it by themselves} do not constitute the full set of structure constants of
any Lie algebra, since they do not satisfy a Jacobi identity. The 
latter result itself is of course well-known.

We draw attention also to \cite{dWn}, which derives identities 
for the torsion tensor that parallelises the sphere $S^7$, this tensor being 
a constant multiple of the one that specifies the octonionic product law. In
particular, one of these identities is equivalent to (\ref{A1.3}).

\section{An equivalent $7 \times 7$ matrix representation}
\subsection{The matrices $H_i$ and $C_a$}

The representation of the matrices $x_i, z_a$ and $y_\al$ that we have been 
lead to above is the one that emerges naturally from the Cartan-Weyl 
representation of $g_2$ as a subalgebra of $b_3$. This enabled the 
definition (\ref{10.2}) -- (\ref{10.7}) of $g_2$ invariant tensors. In virtue
of the conjugation properties (\ref{10.1}) the representation is probably
not the most
convenient one to use in deriving the identities involving these tensors.
Thus we wish to pass to an equivalent basis in which the $x_i$ and $z_a$ are
replaced by antisymmetric matrices, and the $y_\al$ by symmetric ones. In fact
the required basis is already to hand.

To see this we note first that the Lie bracket relation of (\ref{10.9B})
provides us with a set of $14$ hermitian antisymmetric matrices $H_i$ for
the $7 \times 7$ representation $\ga$ of $g_2$ via the definition
\be \label{11.1}
(H_i)_{ab} \; = \; ih_{iab} \qd , \e
since the Jacobi identity for $x_i, x_k, z_a$ translates directly into the 
result
\be \label{11.2}
{[} H_i \, , \, H_k {]} \; = ic_{ikl} H_l \qd . \e
There is clearly an equivalence relation of the form
\be \label{11.3}x_i=S H_i S^{-1} \qd , \qd  \det S \neq 0 \qd , \e
so that we get
\be \label{11.4}
{\rm tr}\; H_i=0 \qd , \qd {\rm tr}\; (H_i \, H_j)=\da_{ij} \qd . \e
This last result translates into the identity
\be \label{11.51}
h_{iab} \, h_{jab} = 2\da_{ab} \qd . \e
Next we examine the Jacobi identity for $x_i, z_a, z_b$. 
Using the two relations  (\ref{10.9A}) and (\ref{10.9B}), we are lead by 
linear dependence to
two relations, one of which reproduces (\ref{11.2}). The other gives
\be \label{11.5}
{[} H_i \, , \, C_a {]} \; = ih_{iab} C_b  \qd , \e
where we have defined a set of $7 \times 7$ antisymmetric matrices $C$ via
\be \label{11.6}
(C_a)_{bc} \; = \; +\eta ic_{abc} \qd , \e
where $\eta$ is a constant that remains to be determined. 
If we normalise $C_a$ so that 
\be \label{11.100}
{\rm tr}\, (C_a \, C_b)=2\da_{ab} \qd , \e
then (\ref{10.10}) and (\ref{A1.1}) give $\eta^2=1$, and it turns out below 
that $\eta=1$.
Eq. (\ref{11.5}) implies that the $C_a$ transform according to the $7$ 
dimensional representation of $g_2$. Further conjugation by $S$ of the Lie 
algebra relation of (\ref{10.9B}) for  gives rise to
(\ref{11.5}) with the identification
\be \label{11.7}
z_a \; = \; S C_a S^{-1} \qd . \e
In other words conjugation with $S$ carries the basis $x_i\; , \; z_a$
of $b_3$ into equivalence with the basis of antisymmetric matrices $H_i \; ,
\; C_a$.

Returning to a topic broached at the end of Sec. 3, we recall that the 
$c_{abc}$ are proportional to the third rank antisymmetric tensor of the
multiplication law of octonions. They have entered first as the structure
constants that reflect the non-symmetric nature of $g_2$ as a 
subalgebra of $b_3$. Second they determine the elements of the matrices $C$
which enter the version
\be \label{11.8}
{[} C_a \, , \, C_b {]} \; = ih_{iab} H_i +ic_{abc} C_c \qd , \e
of the previous statement, {\it i.e.} (\ref{10.9A}), which arises by
conjugation of it by $S$. Use of (\ref{11.8}) and (\ref{11.100}) shows that
\be \label{11.9}
{\rm tr} (C_a C_b C_c) = ic_{abc} \qd , \e
and comparison of this with (\ref{A1.4}) requires the choice $\eta=1$
noted above. Also 
(\ref{11.100}) now translates into the identity
\be \label{11.10}
c_{aef} \, c_{bef} =2\da_{ab} \qd . \e

\subsection{The matrices $Y_\al$}

To get at the matrices generated by conjugation with $S$ of the matrices 
$y_\al$
we set out from the consequence of (\ref{10.6}) 
\be \label{12.1}
{[} x_i \, , \, y_\al{]} \; = i\phi_{i\al \ba} y_{\al \ba}   \qd . \e
obtaining
\be \label{12.2}
{[} H_i \, , \, Y_\al {]} \; = i\phi_{i\al \ba} Y_\ba  \qd , \e
where 
\be \label{12.3} 
Y_\al \; = S (Y_\al) S^{-1} \qd .\e
Also the Jacobi identity containing ${[} \{ z_a \, , z_b \} , y_\al {]}$ 
leads to
two relations. The one we want, from the $x_i$ term, is of the expected form
(\ref{12.2}) if we set
\be \label{12.4}
(Y_\al)_{ab}= \eps d_{ab\al} \qd , \e
where $\eps$ is a constant. To determine $\eps$, we use (\ref{10.3}), the
equivalence relations (\ref{11.3}) and (\ref{11.7}), and (\ref{10.10}) to 
obtain
\be \label{12.5}
d_{ab\al}={\rm tr}\, (z_a z_b y_\al)={\rm tr}\, (C_a C_b Y_\al)=
-\fract{1}{3} \, \psi_{aef} \psi_{fbg} \, \eps d_{ge\al} \qd . \e
Since $Y_\al$ is traceless and symmetric, only one term of (\ref{A1.3})
contributes to (\ref{12.5}). Hence we find
\be \label{12.6}
(Y_\al)_{ab}= -3 \, d_{ab\al} \qd . \e
We also, from equivalence, have 
\be \label{12.7}
{\rm tr}\, (Y_\al \; Y_\ba )= 2\da_{\al \ba} \qd ,\e
and hence
\be \label{12.8}
d_{ab\al} \, d_{ab\ba}=\fract{2}{9} \, \da_{\al \ba} \qd .\e

\subsection{Completeness considerations}

We are in a position to use as a set of $48$ lambda matrices of 
$a_6=su(7)$ the matrices
\be \label{13.1} 
\la_A=\{ H_i \, , \, i \in \{1, \cdots , 14 \} \; ; \; C_a
\, , \, a \in \{1, \cdots , 7\} \; ; \; Y_\al \, , \, 
\, , \, \al \in \{1, \cdots , 27 \} \qd . \e
They are linearly independent, hermitian and traceless, and have been 
normalised in agreement with 
\be \label{13.2}
{\rm tr}\, (\la_A \, \la_B)= 2\da_{AB} \qd . \e
In addition, we have the symmety properties
\be \label{13.3}
H_i=-H_i{}^T \qd , \qd C_a =-C_a{}^T \qd , \qd Y_\al=+Y_\al^T \qd . \e

The completeness identities for $b_3$ and $a_6$ can be written down directly.
For $b_3$ we have
\be \label{13.4}
(H_i)_{ab} \, (H_i)_{cd}+ (C_e)_{ab} \, (C_e)_{cd}
= \da_{ad} \da_{bc}-\da_{ac} \da_{bd} \qd , \e
and, with the aid of the well-known completeness result for $a_6,$ we obtain 
\be \label{13.5}
(Y_\al)_{ab} \, (Y_\al)_{cd}
= -\fract{2}{7} \da_{ab} \da_{cd}+\da_{ad} \da_{bc} +\da_{ac} \da_{bd} \qd . \e
Also (\ref{13.4}) may be arranged as a completeness result for $g_2$ because
of (\ref{10.10}). This gives
\be \label{13.6}
(H_i)_{ab} \, (H_i)_{cd}= \da_{ad} \da_{bc}-\da_{ac} \da_{bd}
+\fract{1}{3} \psi_{eab}\, \psi_{ecd} \qd . \e

Finally by conjugating with $S$ the completeness result for the basis 
$x_\mu= \{x_i \; , \; z_a \} $ for $b_3$, namely
\be \label{13.7}
(x_i)_{ab} \, (x_i)_{cd}+ (z_e)_{ab} \, (z_e)_{cd}
= \da_{ad} \da_{bc}-M_{ac} M_{bd} \qd , \e
and comparing with (\ref{13.3}), one may use Schur's lemma to deduce
\be \label{13.8}
S^T M S = f \, I \qd , \e
where $f$ is a constant that has not been determined.

\section{Bilinear tensor identities}

We begin with a listing of the most important of these `two-tensor' identities
\bea 
c_{ijk} c_{ijl} & = & 8\da_{kl} \label{20.1} \\
h_{iab} h_{jab} & = & 2\da_{ij} \label{20.2} \\
h_{iab} h_{iac} & = & 4\da_{bc} \label{20.3} \\
d_{ij\al} d_{ij\ba} & = & \fract{32}{9} \da_{\al \ba} \label{20.4} \\
d_{ij\al} d_{ik\al} & = & \fract{48}{7} \da_{jk} \label{20.5} \\
c_{abc} c_{abd} & = & 2\da_{cd} \label{20.6} \\
d_{ab\al} d_{ab\ba} & = & \fract{2}{9} \da_{\al \ba} \label{20.7} \\
d_{ab\al} d_{ac\al} & = & \fract{6}{7} \da_{bc} \label{20.8} \\
\phi_{i\al \ba} \phi_{j\al \ba} & = & 18\da_{ij} \label{20.9} \\
\phi_{i\al \ba} \phi_{i\ga \ba} & = & \fract{28}{3}\da_{\al \ga} 
\label{20.10} \\
d_{\al \ba \ga} d_{\al \ba \da} & = & \fract{110}{7} \da_{\ga \da } 
\label{20.11} \\
d_{ia\al} d_{ia\ba} & = & \fract{28}{9} \da_{\al \ba} \label{20.12} \qd . \ea

There are various ways of establishing these results, most of which 
have been confirmed using MAPLE. A systematic approach based on definition of
tensors as traces, followed by use of completeness, is illustrated
below after the collection of a few lemmas. We have, of course, 
found (\ref{20.2}), (\ref{20.6})  and 
(\ref{20.7}) already. Also,  
it is obvious that (\ref{20.3}),
(\ref{20.5}), (\ref{20.8}) and (\ref{20.10}), are easy consequences of their
predecessors; (\ref{20.12}) likewise implies two further related results.

Recalling the definitions
\bea
(ad_i)_{jk} & = & -ic_{ijk} \label{20.20} \\
(H_i)_{ab} & = & -ih_{iab} \label{20.21} \\
(C_a)_{bc} & = & ic_{abc} \label{20.22} \\
(\Phi_i)_{a\al \ba} & = & -i\phi_{i\al \ba}  \label{20.23} \\ 
(Y_\al)_{ab} & = & -3d_{ab\al} \qd , \label{20.24} \ea
we see that the above results can be recast in the useful forms (see Sec. 6.1)
\bea
(ad_i) \, (ad_i) & = & 8 \label{20.30} \\
(H_i) \, (H_i) =x_i x_i & = & 4\label{20.31} \\
(C_a) \, (C_a) =z_a z_a & = & 2 \label{20.32} \\
(\Phi_i) \, (\Phi_i) & = & \fract{28}{3}  \label{20.33} \\
(Y_\al) \, (Y_\al) =  (y_\al) \, (y_\al) & = & \fract{54}{7}  \qd . 
\label{20.34} \ea

\subsection{Some lemmas involving the matrices $H_i \, , \, C_a$ and $Y_\al$}

We begin with a listing
\bea
H_i H_j H_i & = & 0 \label{21.1} \\
H_i C_a H_i & = & 2C_a \label{21.2} \\
H_i Y_\al H_i & = & -\fract{2}{3} Y_\al \label{21.3} \\
C_a H_i C_a  & = & H_i  \label{21.4} \\
C_a C_b C_a  & = & -C_b \label{21.5} \\
C_a Y_\al C_a  & = &   -\fract{1}{3} Y_\al \label{21.6} \\
Y_\al Y_\ba Y_\al & = &   \fract{5}{7} Y_\ba \label{21.7} \\
Y_\al H_i Y_\al & = &   -\fract{9}{7} H_i \label{21.8} \\
Y_\al C_b Y_\al & = &   -\fract{9}{7} C_b  \label{21.9} \qd . \ea

To prove these we first write down these consequences of the completeness 
relation (\ref{13.4})
\bea 
H_i H_k H_i +C_e H_k C_e  & = & H_k  \label{21.21} \\
H_i C_a H_i +C_e C_a C_e  & = & -C_a  \label{21.22} \\
H_i Y_\al H_i +C_e Y_\al C_e  & = &  -Y_\al \qd . \label{21.23} \ea
Also the total antisymmetry of the $c_{abc}$ allows us to make these 
rearrangements
\bea 
(C_e H_k C_e)_{cd} & = & {\rm tr} \, (C_d C_c H_k) =(H_k)_{cd} \nonumber \\
(C_e C_a C_e)_{cd} & = & {\rm tr} \, (C_d C_a C_c)  =-(C_a)_{cd} \nonumber \\
(C_e Y_\al C_e)_{cd}  & = & {\rm tr} \, (C_d C_c Y_\al)=
 -\fract{1}{3} \, (Y_{\al})_{cd} \qd . \label{21.24} \ea
Hence all the first six results listed follow. The last three are easy consequences of ({\ref{13.5}).
 
\subsection{Completion of the proofs of bilinear identities}

To prove (\ref{20.4}), we set out from
\be \label{22.1}
d_{ij\al} \,d_{ij\ba} = {\rm tr} \, (H_i H_j Y_\al) \,  
{\rm tr} \, (H_i H_j Y_\ba) \qd , \e
and use completeness relation (\ref{13.5}), followed by the results 
(\ref{20.31}) and (\ref{21.1}). This gives (\ref{20.4}).
Eqs. (\ref{20.10}) and (\ref{20.11}) follow similarly, and indeed 
(\ref{20.8}), and hence (\ref{20.7}) can be confirmed.

Turning finally to (\ref{20.1}), we use the completeness relation 
(\ref{13.4}) to derive
\be \label{22.2}
 c_{ijk} \, c_{ijl} = -{\rm tr} \, (H_i H_j H_k) \,  
{\rm tr} \, (H_i H_j H_l) =-{\rm tr} \, \bigl( H_j H_k H_j H_l
+H_j H_k H_l H_j  \bigr)
\qd . \e Here the $CC$ term of (\ref{13.4}) has not given any contribution
because $ {\rm tr} \, (C_e H_j H_k)=0$, a result which reflects the closure
properties of the $g_2$ matrices $H_i$. The results
(\ref{20.31}) and (\ref{21.1}) now lead to (\ref{20.1}).

\section{Casimir operators, projectors and characteristic equations}
\subsection{The quadratic Casimir operator}

The irreducible representation of $g_2$ of highest weight $(\la , \mu)$ is well
known , {\it e.g.} \cite{slan}, to have dimension given by
\be \label{25.1}
120 \, \dim \,(\la , \mu)= (\la +1)(\mu+1)(\la + \mu +2)
(2\la + \mu +3)(3\la + \mu +4)(3\la +2 \mu +5) \qd . \e
If $X_i \mapsto D_i$, where these the matrices of $(\la , \mu)$, then the 
quadratic  Casimir operator
\be \label{25.2} \cl{C}^{(2)} =X_i \, X_i \mapsto D_i \, D_i \qd , \e
of $g_2$ has the eigenvalue
\be \label{25.3} 
c^{(2)}(\la , \mu)=\la^2+\fract{1}{3}\mu^2+\la \mu+ 3\la +\fract{5}{3} \mu 
\qd , \e for $(\la , \mu)$.
To within an overall normalisation constant fixed by reference to the 
defining representation of $g_2$, this agrees with the result of \cite{ok1}.
See also \cite{sp}, especially the striking and amusing section III.

In our work so far the representations 
\bea 
(0,1) & = & 7 \, , \quad X_i \mapsto x_i \qd {\rm or} \qd X_i \mapsto H_i
\nonumber \\
(1,0) & = & 14= ad \, , \quad X_i \mapsto (ad_i) \qd , \qd 
(ad_i)_{jk} =-i c_{ijk} \nonumber \\
(0,2) & = & 27  \, , \quad X_i \mapsto \Phi_i \qd , (\Phi_{i})_{\al \ba}=
-i\phi_{i\al \ba}\qd . \label{25.4} \ea
 have occurred. For these (\ref{25.3}) requires the eigenvalues 
$4, 8$ and $\fract{28}{3}$ respectively, and we can see that (\ref{20.30}) --
(\ref{20.33}) are in agreement with this.

\subsection{Projectors}

We consider first the Clebsch-Gordan series
\bea
 \begin{array}{ccccc}
7 \otimes 7 & \equiv &  (1 +27) & + & (7+14) \\
(0,1) \otimes (0,1) & \equiv &  \underbrace{(0,0)+(0,2)}_S & + & 
\underbrace{(0,1)+(1,0}_A) 
\end{array}  \qd . \label{26.1} \ea
Suppose $A_a, B_a, \qd a \in \{1, \cdots , 7 \}$ transform according to the 
defining representation $\ga = 7 = (0,1)$ of $g_2$. The tensors which 
transform according to the symmetric part of $7 \otimes 7$ are given by
\bea
T^{(1)}{}_{ab} & = & \fract{1}{7}\da_{ab} A_c B_c \nonumber \\
 & = & \fract{1}{7} \da_{ab} \da_{cd} A_c B_d \label{26.2} \\
T^{(27)}{}_{ab} & = & \fract{1}{2} (A_a B_b+ A_b B_a)-
\fract{1}{7}\da_{ab} A_c B_c \nonumber  \\
 & = & \bigl[ \fract{1}{2} (\da_{ac} \da_{bd} +\da_{ad} \da_{bc} )-
\fract{1}{7}\da_{ab} \da_{cd} \bigr] A_c B_d \qd . \label{26.3} \ea
Similarly for the antisymmetric part, we have
\bea
T^{(7)}{}_{ab} & = & \fract{1}{2} c_{abe} c_{cde} A_c B_d \label{26.4} \\
T^{(14)}{}_{ab} & = & \fract{1}{2} (A_a B_b-A_b B_a)
-\fract{1}{2} c_{abe} c_{cde} A_c B_d \label{26.5} \\
 & = & \bigl[ \fract{1}{2} (\da_{ac} \da_{bd} -\da_{ad} \da_{bc})
-\fract{1}{2} c_{abe} c_{cde} \bigr]  A_c B_d \qd . \label{26.6}
\ea

Now
\be \label{26.7}
T^{(R)}{}_{ab}=P^{(R)}{}_{ab,cd}A_c B_d \qd , \e
defines a set of orthogonal projectors onto the representations of $g_2$ 
contained in the reduction (\ref{26.1}) of $7 \otimes 7$. 

The above used the result (\ref{20.6}). Also we see that (\ref{13.5}) implies
\be \label{26.8} 
(Y_\al \otimes Y_\al){}_{ac,bd} = 
(Y_\al){}_{ab} (Y_\al){}_{cd}=2 P^{(27)}{}_{ab,cd} \qd . \e
Also 
\be \label{26.9} 
(H_i){}_{ab} (H_i){}_{cd}+(C_e){}_{ab} (C_e){}_{cd}=
2 P^{(A)}{}_{ab,cd} =(\da_{ac} \da_{bd} -\da_{ad} \da_{bc}) \qd , \e
where $P^{(A)}=P^{(7)}+P^{(14)}$ is the projector onto the antisymmetric 
tensor subspace.
In fact, more explicitly
\bea
(C_e){}_{ab} (C_e){}_{cd} & = & -c_{eab} \, c_{ecd} =-2 P^{(7)}{}_{ab,cd} 
\label{26.10} \\
(H_i){}_{ab} (H_i){}_{cd} & = &  (\da_{ac} \da_{bd} -\da_{ad} \da_{bc})
-c_{eab} \, c_{ecd} = 2P^{(14)}{}_{ab,cd} \qd . \label{26.11} \ea

We consider next the Clebsch-Gordan series
\bea
\begin{array}{ccccc}
14 \otimes 14 & \equiv & (1 +27+77) & + & (14+77^\prime ) \\
(1,0) \otimes (1,0) & \equiv & \underbrace{(0,0)+(0,2)+(2,0)}_S & + & 
\underbrace{(1,0)+(0,3}_A) 
\end{array}  \qd . \label{26.12} \ea
Suppose $A_i, B_i, \qd i \in \{1, \cdots , 14 \}$ transform according to the 
adjoint representation $ad = 14 = (1,0)$ of $g_2$. The tensors which 
transform according to the symmetric part of $14 \otimes 14$ are given by
\bea
T^{(1)}{}_{ij} & = & \fract{1}{14} \da_{ij} \da_{kl} A_k B_l \label{26.13} \\
T^{(27)}{}_{ij} & = & \fract{9}{32} d_{ij\al} d_{kl\al} 
 A_k B_l \label{26.14} \\
T^{(77)}{}_{ij} & = &
\bigl[
\fract{1}{2}(\da_{ik} \da_{jl}+\da_{il} \da_{jk})-  \fract{1}{14 } \da_{ij} 
\da_{kl} -\fract{9}{32} d_{ij\al} d_{kl\al} \bigr]  
 A_k B_l \qd .\label{26.15} \ea
The result (\ref{20.4}) has been used here. Similarly, with the aid of 
(\ref{20.1}), we get, for the antisymmetric part, 
\bea 
T^{(14)}{}_{ij} & = & \fract{1}{8} c_{ijp} c_{klp} A_k B_l \label{26.16} \\
T^{(77^{\prime})}{}_{ij} & = & \bigl[ \fract{1}{2} 
(\da_{ik} \da_{jl} - \da_{il} \da_{jk})
-\fract{1}{8} c_{ijp} c_{klp} \bigr]  A_k B_l \qd . \label{26.17}
\ea

Now
\be \label{26.18}
T^{(R)}{}_{ij}=P^{(R)}{}_{ij,kl}A_k B_l \qd , \e
defines a set of orthogonal projectors onto the representations of $g_2$ 
contained in the reduction (\ref{26.12}) of $14 \otimes 14$. 

One aspect of (\ref{26.12}) is worth clarifying. How did we assign $77=(2,0)$
to the symmetric part of $14 \times 14$, and $77^\prime = (0,3)$ to the 
antisymmetric part? One way, which follows a general argument given on P3196 of
\cite{mpf}, is indicated at the end of Sec. 6.3. 
 
There is a very interesting discussion of projectors for $g_2$ in \cite{bdfl}.

\subsection{Characteristic equations}

As indicated a long time ago for $su(n)$ \cite{msw}, and emphasised in  
\cite{mpf}, 
there are two classes of basic identities for the invariant
tensors of any Lie algebra -- first class  identities that stem directly from 
Jacobi identities, and those of the second class. The latter depend on the 
use of the characteristic equation of the Lie algebra usually for the 
defining representation $\cl{V}$.
The difficulty of finding them increases with the 
dimension of the Lie algebra or of $\cl{V}$. 
\cite{mpf} 
presents a systematic
account with lots of results including many for $g_2$. It is useful to 
present a slightly modified version of the latter discussion. Some of the 
intermediate equations of the discussion may here appear slightly different 
from their counterparts in \cite{mpf}, because normalisations used there were
fixed in a way that allowed uniform programming for all Lie algebras, whereas
here we do what seems most convenient for $g_2$ itself.

Let $ad_1$ and $ad_2$ denote adjoint representations of $g_2$
acting in vector spaces $\cl{V}_1$ and  $\cl{V}_2$, of dimension $14$. The
Clebsch-Gordan series for $14 \otimes 14$ is displayed as (\ref{26.12}).
Let $X_i=X_{i1}+X_{i2}$ denote the `total' $g_2$ generators, acting in
$\cl{V}_1 \otimes \cl{V}_2$. Then 
\be \label{27.1}
X_i \, X_i=X_{i1}\, X_{i1}+X_{i2}\, X_{i2}+ 2\La \qd , \qd 
\La=X_{i1}\, X_{i2} \qd , \e
implies that $\La$ has the same eigenspaces as does $X_i \, X_i$, and 
eigenvalues $-8, -\fract{10}{3}, 2, -4, 0$ for the representations
$1, 27, 77 ,14, 77^{\prime}$ respectively. To use this information, we begin 
by recalling a well-known formula 

If a hermitian operator $A$ has eigenvalues and eigenvectors given by
\be \label{27.3} 
A |a_i\rangle =a_i |a_i\rangle \, , \qd a \in \{ 1, \cdots ,p 
\} \qd , \e
then the projector onto its  $i$-th eigenspace is
\be \label{27.4}
P_i=\prod_{k \neq i} \frac{A-a_k I}{a_i-a_k} \qd , \e
where $I$ is the unit operator in the vector space spanned by the 
$|a_i\rangle$.
Clearly
\be \label{27.5}
(A-a_i I) \, P_i=0 \qd ,\e
with no sum on $i$ implied,
is, for each $i$, a, possibly reduced, version of the characterictic equation 
sought. Now we know that $\La$ is a linear combination of the projectors onto 
the eigenspaces of $X_i \, X_i$. Further $\La \, I_S$, where $(I_S)_{ij,kl}=
\fract{1}{2}(\da_{ik} \da_{jl}+\da_{il} \da_{jk})$ is the unit of the
symmetric subspace, is a linear combination of the symmetric projectors for the
representations $1, 27, 77$. For these, (\ref{27.5}) implies
\bea
0 & = & (\La +8) \, I_S P^{(1)} \label{27.6} \\
0 & = & (\La +\fract{10}{3} ) \, I_S P^{(27)} \label{27.7} \\
0 & = & (\La -2) \, I_S P^{(77)} \qd . \label{27.8} \ea
We could well have left out $I_S$ from these equations.
Since
\be \label{27.9}
P^{(1)}+ P^{(27)}+ P^{(77)}= I_S \qd , \e
the last four equations lead easily to 
\bea 
0 & = & \La I_S+8 P^{(1)}+ \fract{10}{3}  P^{(27)}- 2 P^{(77)} \nonumber \\
0 & = & \La I_S+ 10 P^{(1)}+ \fract{16}{3}  P^{(27)}- 2 I_S \qd. \label{27.10} \ea

Since
\be \label{27.11} (\La)_{rs,ij}= -c_{pri} c_{psj} \qd ,  \e
and since we get the explicit expressions for the projectors from
(\ref{26.13}) and (\ref{26.14}), we are directly lead to the important 
result
\be \label{27.12}
3\, d_{rs\al}\, d_{kl\al}=c_{prk} c_{psl}+c_{prl} c_{psk}
+2(\da_{rk} \da_{sl}+\da_{rl} \da_{sk}) -\fract{10}{7}\da_{rs} \da_{kl}
\qd . \e
This is a second class result because its derivation employs a 
characteristic equation. It is also 
an analogue of eq. (2.23) of \cite{msw}, and is found in \cite{mpf} as
eq. (4.32b) of that paper. 
We note some immediate consequences. With the aid of the ordinary Jacobi 
identity,
applied twice, once to treat each of the first two terms of the right side of
(\ref{27.12}), we deduce a further important result
\be \label{27.13}
d_{(ij}{}^\al \, d_{k)l\al} \; = \; \fract{6}{7} \da_{(ij} \, \da_{k)l} \qd .
\e 
Here the round brackets indicate 
symmetrisation of unit weight over the indices enclosed; sometimes indices
are raised in a manner with no metric significance in order to exempt them from
symmetrisation. The same comment will later apply to
antisymmetrisation and square brackets. Also, we may apply $d_{is\ba}$ to
(\ref{27.12}) and deduce the identity
\be \label{27.14} 
c_{pri} c_{psj} d_{ij\al}=\fract{10}{3} \,d_{rs\al}
\qd . \e

Because it leads us to important results, we consider the what the 
above analysis can tell us if we wish to avoid any mention of the tensors
$d_{ij\al}$. As in 
\cite{mpf}, 
we apply $\La$ to (\ref{27.10}), and 
then use this equation again to eliminate $P^{(27)}$. Hence we find
\be \label{27.15}
0 = {\La}^2 \, I_S+ \fract{4}{3} \La \, I_S-\fract{140}{3} P^{(1)}- 
\fract{20}{3}I_S \qd . \e 

It is convenient to add to this the equation 
\be \label{27.16} \La^2 I_A +4\La I_A=0 \qd . \e
To establish (\ref{27.16}), we make use of 
\be \label{27.17}
(\La  +4) \, I_A  P^{(14)}=0  \qd , \e
which follows from (\ref{27.5}), and
\be \label{27.18}
\La I_A +4 P^{(14)}=0 \qd , \e
which coincides with the Jacobi identity of $g_2$. 
This leads us to (\ref{27.16}) and to the desired formula 
\be \label{27.19}
0 = \La^2+\fract{4}{3} \La \, I_S +4 \La \, I_A-\fract{140}{3} P^{(1)}
-\fract{20}{3} I_S \qd , \e and hence to  
\be \label{27.20}
3c_{jmr} c_{knr} c_{mps}c_{nqs} = 10(\da_{jp} \da_{kq}+\da_{jq} \da_{kp}
+\da_{jk} \da_{pq})
+8 c_{jpr} c_{kqr} - 4 c_{jqr} c_{kpr} \qd. \e
This can be written out as a formula for
${\rm tr}\, ad_j \; ad_k \; ad_p \; ad_q$, and implies 
\be \label{27.21}
{\rm tr}\, ad_{(j} \; ad_k \; ad_p \;  ad_{q)} = 10 \da_{(jk} \da_{pq)} 
\qd . \e
 
As a  further consequence of (\ref{27.16}), we note $\La I_A$ has eigenvalues
$-4$ and $0$, corresponding to the representations $14=(1,0)$ and
$77^\prime = (0,3)$, thereby confirming that $(0,3)$ 
rather than $77=(2,0)$ is a
constituent of the antisymmetric part of $14 \otimes 14$.

\subsection{On the non-primitive quartic Casimir operator}

First we quote the evaluation of a trace
\be \label{28.1}
{\rm tr}\, x_{(i} x_j x_k x_{l)} \equiv 
{\rm tr}\, H_{(i} H_j H_k H_{l)} =\da_{(ij}\, \da_{kl)} \qd . \e

Proof:
A routine calculation based on two uses of (\ref{10.2}) and evaluation of 
elementary traces yields the answer
\be \label{28.2}
\fract{4}{7} \da_{(ij} \, \da_{kl)}+ \fract{1}{2} \, 
d_{(ij}{}_\al \, d_{k)l\al} \qd , \e
and then (\ref{27.13}) is used.

Second we define $A= A_i x_i$, $A_i \in {\BB R}$, and show that
\be \label{28.3} 
{\rm tr}\, A^4 = (\fract{1}{2}\,{\rm tr}\, A^2)^2 \qd . \e
This trivial consequence of (\ref{28.1}) indicates the absence 
(see
 \cite{ok1} 
for an early proof) of a primitive 
quartic Casimir operator for $g_2$. It reflects the well-known 
fact that there are just two  
primitive Casimir operators of $g_2$, which have orders $2$ and $6$. 

Third, we shall prove a stronger result regarding the non-primitive nature of
the quartic Casimir of $g_2$. We introduce $X=x_i X_i$ where the $x_i$ are the
matrices of the $7 \times 7$ defining representation, and the $X_i$ are the
abstract generators. Then define the quartic Casimir operator of $g_2$ by 
means of
\be \label{29.1}
\cl{C}^{(4)}={\rm tr}\, X^4 ={\rm tr}\, (x_i x_j x_k x_l) X_i X_j X_k X_l
\qd . \e
To evaluate this correctly requires taking full account of the Lie algebra 
relations
\be \label{29.2}
{[} X_i \, , \, X_j {]}=ic_{ijk} X_k \qd . \e
It is routine to obtain
\be \label{29.3}
{\rm tr}\, (x_i x_j x_k x_l)={\rm tr}\, (x_{(i} x_j x_k x_{l)})
-\fract{1}{3} (c_{klt}\, c_{ijt} - c_{ilt} \, c_{jkt}) \qd . \e
Using this result and (\ref{28.1}) enables the contribution from the first
term of (\ref{29.3}) to (\ref{29.1}) to be evaluated. The contribution from 
the other terms depends only on the use of (\ref{29.2}), and of the identity
(\ref{20.1}). One is lead then to the answer
\be \label{29.4}
\cl{C}^{(4)}=(\cl{C}^{(2)}){}^2+ \fract{28}{3} \cl{C}^{(2)} \qd . \e
Again this result agrees with that given in \cite{ok1}. Okubo did not give
details of his proof, indicating just that it was `involved'. 
The present proof
depends on the introduction of a full range of invariant tensors and on gaining
full control of their properties.

\section{Trilinear tensor identities}

We first remark that the product laws (\ref{10.2}) -- (\ref{10.7})
give rise easily to one family of `three-tensor' identities. For example
\be \label{35.1}
c_{ijk}= -i {\rm tr} \, x_i x_j x_k= -i{\rm tr} \, H_i H_j H_k 
=h_{iab} \, h_{jbc} \, h_{kca} \qd \, \e
upon use also of (\ref{11.3}) and \ref{11.1}). Such expressions also, as noted,
account for the appearance in later product laws of tensors defined in earlier
ones. We quote two more results from this family
\bea
c_{abc} & = & c_{eaf} c_{fbg} c_{gce}  \label{35.2} \\
d_{\al \ba \ga}  & = & -27 d_{ab\al} d_{bc\ba} d_{ca\ga} \qd . \label{35.3} \ea
Eq. (\ref{35.2}) coincides with (\ref{21.5}).

Next we give a listing of three tensor identities that do not arise in the 
same way as those just noted.
\bea 
c_{piq} c_{qjr} c_{rkp} & = & -4c_{ijk} \label{35.4} \\
d_{jk\al} d_{li\al} c_{jlq} & = & \fract{20}{7} c_{kiq} \label{35.5} \\
c_{pri} c_{psj} d_{ij\al}  & = & \fract{10}{3} d_{rs\al} \label{35.6} \\
d_{ij\al}  d_{jk\ba} d_{\al \ba \ga} & = & \fract{22}{21} d_{ik\ga } 
\label{35.7} \\
d_{pq\al}  d_{pi\ba} d_{qj\ba}  & = & -\fract{58}{63} d_{ij\al} \label{35.8} \\
d_{\la \mu \al} d_{\mu \nu \ba} d_{\nu \la \ga } & = & \fract{53}{7}   
d_{\al \ba \ga} \qd . \label{35.9} \ea
We turn to the proof of these results. All have been
confirmed using MAPLE.
By  a method that applies to any Lie algebra, we see that (\ref{35.4}) 
follows directly from the 
Jacobi identity for $g_2$, upon use of (\ref{20.1}). To prove (\ref{35.5}), we reduce
\be \label{35.11}
{[} {[} x_i \, , \,x_j {]} ,\, x_k {]}= \{ \{ x_j \, , \, x_k \} , \, x_i \}
- \{ \{ x_k \, , \, x_i \} , \, x_j \} \qd , \e
and reach an identity of the first class
\be \label{35.12}
c_{ijp} c_{klp}= \fract{8}{7} (\da_{ik} \da_{jl}-\da_{jk} \da_{il})
+(d_{ik\al}d_{jl\al}- d_{jk\al} d_{il\al}) \qd .\e
Now apply $c_{jlq}$ to (\ref{35.12}). The result simplifies, with the aid of 
(\ref{35.4}) and (\ref{20.1}) to yield (\ref{35.5}).

We have obtained (\ref{35.6}) already as (\ref{27.14}).
Next we apply $d_{ik\ba}$ to (\ref{35.12}) getting
\be \label{35.51} 
c_{pri} c_{psj} d_{ij\ba}+d_{pq\ba} d_{pr\al} d_{qs\al} =\fract{152}{63}
d_{rs\ba} \qd . \e 
Now (\ref{35.6}) implies (\ref{35.8}).

The results (\ref{35.7}) and (\ref{35.9}) are easy to prove because the simple
completeness relation (\ref{13.5}) can be used. For example
\be \label{35.13}
d_{ij\al} d_{jk\ba} d_{\al \ba \ga}=(H_iH_j)_{ba} (H_j H_k)_{dc} 
(Y_{\al}){}_{ab} (Y_{\ba}){}_{cd} (Y_{\al}){}_{ef} (Y_{\ba}){}_{fg}
(Y_{\ga}){}_{ge} \qd . \e
Two applications of (\ref{13.5}) now allow the proof of (\ref{35.7}) to be
completed. The results (\ref{11.4}), (\ref{20.31}), (\ref{20.1}),
(\ref{13.5}) and $d_{ij\al}= {\rm tr}\, H_i H_j Y_\al$
are all used in the process.

Finally we note the absence of any simple result like those just proved
for the quantity
\be \label{35.14}
d_{ij\al}  d_{jk\ba} d_{ki\ga} \qd . \e
It would be wrong to suppose this is a constant multiple of $d_{\al \ba \ga}$. 
No such result exists, and MAPLE rejects such a conjecture. The vector space of
third rank tensors totally symmetric in $\al , \ba , \ga$ has been  analysed
fully but we have no good reason to record the details.

\section{Adjoint vectors and invariants}
\subsection{Results}

Let the vector $A_i, \; i \in \{ 1, \cdots , 14\}$ transform under the 
action of $g_2$ according to its adjoint representation. Then the vector 
\be \label{40.1}
d_{ij}{}^\al d_{kl\al} A_j A_k A_l= \fract{6}{7} (A_p A_p) \, A_i \qd , \e
is seen not to be linearly independent of $A_i$. Eq. (\ref{40.1}) holds
because we can freely
put round symmetrisation brackets round the set of indices $j, k, l$, and 
employ
the identity (\ref{27.10}). The result is tantamount to the fact that the
quartic Casimir of $g_2$ is not primitive. It suggests 
\cite{ok1}
that there should be a
second adjoint vector, whose components are quintic in those of $A_i$, since 
we know a primitive sixth order invariant exists. In fact, the required
vector $C_i$  is given by
\be \label{40.2}
C_i=d_{\al \ba \ga} d_{ij}{}^\al d_{kl}{}^\ba d_{pq\ga}
A_j A_k A_l A_p A_q \qd . \e
Here we can use the $A_r$ factors to justify putting round brackets 
around the index set $jklpq$, and then
see that $i$ can also be accommodated correctly within them because of the 
symmetry of the tensorial factors. Thus we write
\be \label{40.3}
C_i=T_{ijklpq} A_j A_k A_l A_p A_q \qd,. \e
where the totally symmetric sixth rank tensor is given by
\be \label{40.4}
T_{ijklpq} = d_{\al \ba \ga} d_{(ij}{}^\al d_{kl}{}^\ba d_{pq)\ga} \qd . \e

Turning next to the construction of invariants or scalars out of the components
of $A_i$, we see that there are quadratic and sixth order invariants
\be \label{40.5}
A_i \, A_i \; , A_i \, C_i \qd , \e
as expected. However there are others. One is $C_i C_i$. Others use the 
$27$-component quantities 
\be \label{40.6} 
B_\al= d_{ij\al} A_i A_j \qd , \qd D_\al =d_{\al \ba \ga} B_\ba B_\ga \qd , \e
to build other scalars  
\be \label{40.7} 
B_\al \, B_\al \; , \; B_\al \, D_\al \; , \; D_\al \, D_\al \qd . \e
None of the additional scalars are primitive. Indeed easily we can see
\be \label{40.8}
B_\al \, D_\al=  A_i \, C_i \; , \; B_\al \, B_\al =\fract{6}{7} 
A_i \, A_i \, A_j \, A_j \qd , \e
leaving $C_i C_i$ and  $D_\al \, D_\al$ to be given explicitly in terms of
primitive invariants.

Also, the tensor $T_{ijklpq}$, while totally symmetric, is not traceless. For 
the optimal construction a sixth order scalar,
we should construct 
\cite{tensors} \cite{dAMcas} \cite{vRSV} 
the traceless totally symmetric tensor
\be \label{40.9}
S_{ijklpq}=T_{ijklpq} -\fract{88}{441} \da_{(ij} \da_{kl} \da_{pq)} \qd . \e
such that
\be \label{40.10}
S_{iiklpq}=0 \qd . \e
To establish (\ref{40.10}), we need to open out the round brackets of the 
definition (\ref{40.4}). Upon putting $j=i$, some terms vanish because of 
$d_{ii\da}=0$. In fact this applies to $3$ out of $15$ distinct terms.
The remainder are all equivalent, and such that (\ref{35.7})
can be applied. This means that
\be \label{40.11}
T_{iiklpq}=\fract{4}{5} \; \fract{22}{21} \; \fract{6}{7} \da_{(kl} \da_{pq)}
\qd , \e in which, at the very end, (\ref{27.13}) has been used. Since
putting $j=i$ in the $\da\da\da$ term gives the same result, the proof is 
done.

\subsection{On the sixth order invariant and non-primitive invariants}

To get some measure of control of the sixth order Casimir operator, and to 
treat explicitly the non-primitivity of scalars such as $C_i C_i$ and 
$D_\al \, D_\al$ of orders $10$ and $8$, we use a basis in which $A_i$ has the
components $(a, b, 0, \cdots ,0)$ so that $A=A_i \, x_i$ is diagonal
\be \label{41.1} 
A={\rm diag} \, ( \sqrt{\fract{2}{3}}a, \sqrt{\fract{1}{3}}a+ 
\sqrt{\fract{1}{2}}b, \sqrt{\fract{1}{3}}a-\sqrt{\fract{1}{2}}b, 0,
 -\sqrt{\fract{1}{3}}a+\sqrt{\fract{1}{2}}b, -\sqrt{\fract{1}{3}}a-
\sqrt{\fract{1}{2}}b, - \sqrt{\fract{2}{3}}a ) \qd .\e
From this we find
\be \label{41.2}
\cl{C}^{(2)} = A_i \, A_i = \fract{1}{2} \, {\rm tr}\, A^2=(a^2+b^2) \; . \e

It is easy to evaluate explicitly the few tensor components needed
to show that  
\bea
B_\al & = & (\sqrt{\fract{2}{3}}\; (a^2-b^2), \sqrt{\fract{2}{3}}2ab, 
\sqrt{\fract{4}{21}} (a^2+b^2), 0, \cdots , 0) \label{41.3} \\
D_\al & = & (\sqrt{\fract{2}{3}}(\fract{22}{21}a^4-4a^2b^2+\fract{2}{7}b^4,
\sqrt{\fract{2}{3}}(-\fract{40}{21}a^3b+\fract{24}{7}ab^3), 
-\fract{4}{7} (\sqrt{\fract{1}{21}}(a^2+b^2)^2, 0, \cdots ,0) \, ,
\nonumber \ea 
where the components $\al=1,2,3$ correspond to
the ordering of (\ref{5.3}) and (\ref{5.5}).
This enables us to verify easily the second result in (\ref{40.8}), and to use
the first one to deduce
\be \label{41.4}
\cl{C}^{(6)} = A_i \, C_i=\fract{88}{441} \, {\cl{C}^{(2)}}{}^3
+\fract{4}{9}\, (a^2-b^2) (a^4-14a^2 b^2+b^4)  \; . \e
 
The fraction in the first term is the same one as we found by independent 
calculations in (\ref{40.9}). Thus if we define the optimal sixth order 
invariant ${\tilde {\cl{C}}^{(6)}}$ that can be built using six copies of the 
vector $A_i$ as
\be \label{41.5}
{\tilde {\cl{C}}{}^{(6)}} =S_{ijklpq} A_i A_j A_k A_l A_p A_q \qd , \e
then its value in terms of $a, b$ is
\be \label{41.6}
{\tilde {\cl{C}}{}^{(6)}} = \fract{4}{9} \, (a^2-b^2) (a^2-4ab+b^2)
(a^2+4ab+b^2) \qd . \e

The data accumulated in this section enables us to show
\be \label{41.7}
D_\al D_\al=\fract{16}{21} \cl{C}^{(2)} \, \cl{C}^{(6)}
+\fract{88}{343} \, {\cl{C}^{(2)}}{}^4 \qd . \e

To treat $C_i \, C_i$, we note that
\be \label{41.8}
C_i =d_{ij\al} a_j D_\al \qd . \e
All the information is therefore at hand to allow us to obtain the result
\be \label{41.9}
C_i \, C_i = \fract{16}{147} {{\cl C}^{(2)}}{}^2 \Bigl( \fract{11}{3}
{\cl C}^{(6)}+ \fract{71}{49 } {{\cl C}^{(2)}}{}^3  \Bigr) \qd , \e
it being easy to see how the factor ${{\cl C}^{(2)}}{}^2$ arises.
MAPLE confirms the above results.

\subsection{Further use of characteristic equations}

Setting $A=A_i \, x_i$, we use the (easily programmable \cite{mpf}) 
methods of Sec. 8.2 to find
\be \label{47.1}  
{\rm tr}\, A^4= \fract{1}{4} ({\rm tr}\, A^2){}^2 \qd , \e
the characteristic polynomial of $A$
\be \label{47.2}
\chi_A(t) = t^7- \fract{1}{2} ({\rm tr}\, A^2) \, t^5 +
\fract{1}{16} ({\rm tr}\, A^2)^2 \, t^3+\Bigl(\fract{1}{96} ({\rm tr}\, A^2)^3-
\fract{1}{6} ({\rm tr}\, A^6)\, \Bigr) \, t \qd , \e
and hence
\bea 
({\rm tr}\, A^8)& = & -\fract{5}{192} ({\rm tr}\, A^2)^4
+ \fract{2}{3} ({\rm tr}\, A^2) \, ({\rm tr}\, A^6) \label{47.3A} \\
({\rm tr}\, A^{10})& = & -\fract{1}{64} ({\rm tr}\, A^2)^5
+ \fract{5}{16} ({\rm tr}\, A^2)^2  \, ({\rm tr}\, A^6) \qd , \label{47.3B} \ea
which agrees with results established in 
\cite{arthur} 
using a different 
method. From these results we can deduce the trace formulas
\bea
{\rm tr}\, \bigl( x_{(i_1} \, x_{i_2} \cdots x_{i_4)} \bigr) & = & 
\da_{(i_1 i_2} \, \da_{i_3 i_4)} \label{47.4} \\
{\rm tr}\, \bigl( x_{(i_1} \, x_{i_2} \cdots x_{i_8)} \bigr) & = & 
=-\fract{5}{12} \da_{(i_1 i_2} \cdots \da_{i_7 i_8)} + \fract{4}{3}
\da_{(i_1 i_2} \, {\rm tr}\, \bigl( x_{i_3} \, x_{i_4} \cdots x_{i_8)} \bigr)
\label{47.5} \\
{\rm tr}\, \bigl( x_{(i_1} \, x_{i_2} \cdots x_{i_{10)}} \bigr) & = & 
-\fract{1}{2} \da_{(i_1 i_2} \cdots \da_{i_9 i_{10})} + \fract{5}{4}
\da_{(i_1 i_2} \, \da_{i_3 i_4} {\rm tr}\, \bigl( x_{i_5} \, x_{i_6} \cdots
 x_{i_{10})} \bigr)
\qd , \label{47.6} \ea
given in terms of the two independent primitive traces
\be \label{47.7} 
{\rm tr}\, ( x_{(i_1} \, x_{i_2)} ) = 2\da_{i_1 i_2} \qd , \qd
{\rm tr}\, \bigl( x_{(i_1} \, x_{i_2} \cdots x_{i_6)} \bigr) \qd . \e
We can relate the latter trace to the tensors used in Sec. 8.1
to define the sixth order Casimir operator of $g_2$. With the help of 
(\ref{10.2}) and (\ref{40.4}), we find
\be \label{47.8}
{\rm tr}\, \bigl( x_{(i_1} \, x_{i_2} \cdots x_{i_6)} \bigr)=
{\rm tr}\, \bigl( x_{((i_1} \, x_{i_2)} \,  x_{(i_3} \, x_{i_4)}\,
 x_{(i_5} \, x_{i_6))} \bigr) =
\fract{26}{49} \da_{(i_1 i_2} \da_{i_3 i_4} \da_{i_5 i_6)}
+\fract{1}{8} T_{i_i \cdots i_6} \qd . \e
 
Turning next to the characteristic equation of the adjoint matrix
$B=A_i \, ad_i$, where $(ad_i)_{jk}=-ic_{ijk}$, we quote from \cite{mpf} the
results 
\bea 
{\rm tr}\, B^2  & = & 4({\rm tr}\, A^2) =8 A_i A_i \label{47.9} \\
{\rm tr}\, B^4  & = & \fract{5}{2} ({\rm tr}\, A^2)^2 \label{47.10} \\
{\rm tr}\, B^6  & = & \fract{15}{4} ({\rm tr}\, A^2)^3
-26 ({\rm tr}\, A^6)  \label{47.11} \\
{\rm tr}\, B^8  & = & \fract{515}{96} ({\rm tr}\, A^2)^4-
\fract{160}{3} ({\rm tr}\, A^2) \, ({\rm tr}\, A^6) \label{47.12} \\
{\rm tr}\, B^{10}  & = &  \fract{431}{64 } ({\rm tr}\, A^2)^5-
\fract{605}{8} ({\rm tr}\, A^2)^2 \, ({\rm tr}\, A^6) 
 \qd , \label{47.13} \ea
noting that the characteristic equation, \cite{mpf} eq. (A44), allows higher 
traces to be computed.
Eq. (\ref{47.10}) is equivalent to (\ref{27.21}). From (\ref{47.11}),
we get
\be \label{47.14}
{\rm tr}\, \bigl( ad_{(i_1} \cdots ad_{i_6)} 
=\fract{794}{49}  \da_{(i_1 i_2} \da_{i_3 i_4} \da_{i_5 i_6)}
-\fract{13}{4} T_{i_i \cdots i_6} \qd ,  \e
and so on.

\end{document}